\documentclass[a4paper,11pt]{article}
\usepackage[utf8]{inputenc}

\usepackage{framed}
\usepackage{mdframed}

\usepackage{fullpage}
\usepackage[T1]{fontenc}

\usepackage[british]{babel}
\usepackage{verbatim}
\usepackage[T1]{fontenc}
\usepackage{lmodern}
\usepackage{lipsum}
\usepackage{booktabs}
\usepackage{caption}
\usepackage{cite}
\usepackage{soul,color}
\usepackage[toc,page]{appendix}

\usepackage{ifpdf}

\usepackage{amsthm, amssymb}
\usepackage[tbtags]{amsmath}
\usepackage{bm}
\usepackage{mathtools}
\usepackage{amstext}
\usepackage{braket}
\usepackage{multirow}

\usepackage[normalem]{ulem}
\usepackage{xcolor}

\newcommand\redsout{\bgroup\markoverwith{\textcolor{red}{\rule[0.5ex]{2pt}{0.4pt}}}\ULon}

\begin{document}
\vspace{5mm}
\vspace{0.5cm}

\def\be{\begin{eqnarray}}
\def\ee{\end{eqnarray}}

\def\ba{\begin{aligned}}
\def\ea{\end{aligned}}

\def\ls{\left[}
\def\rs{\right]}
\def\lc{\left\{}
\def\rc{\right\}}

\def\p{\partial}

\def\S{\Sigma}

\def\s{\sigma}

\def\O{\Omega}

\def\a{\alpha}
\def\b{\beta}
\def\g{\gamma}

\def\ad{{\dot \alpha}}
\def\bd{{\dot \beta}}
\def\gd{{\dot \gamma}}
\newcommand{\ft}[2]{{\textstyle\frac{#1}{#2}}}
\def\ib{{\overline \imath}}
\def\jb{{\overline \jmath}}
\def\Re{\mathop{\rm Re}\nolimits}
\def\Im{\mathop{\rm Im}\nolimits}
\def\trace{\mathop{\rm Tr}\nolimits}
\def\rmi{{ i}}

\def\N{\mathcal{N}}

\newcommand{\SU}{\mathop{\rm SU}}
\newcommand{\SO}{\mathop{\rm SO}}
\newcommand{\U}{\mathop{\rm {}U}}
\newcommand{\USp}{\mathop{\rm {}USp}}
\newcommand{\OSp}{\mathop{\rm {}OSp}}
\newcommand{\Symp}{\mathop{\rm {}Sp}}
\newcommand{\Sl}{\mathop{\rm {}S}\ell }
\newcommand{\Gl}{\mathop{\rm {}G}\ell }
\newcommand{\Spin}{\mathop{\rm {}Spin}}
% algebras
%\newcommand{\so}{\mathfrak{so}}
%  PDF specials
%\newif\ifpdf
%overstruck text
%\newcommand\redsout{\bgroup\markoverwith{\textcolor{red}{\rule[0.5ex]{2pt}{0.4pt}}}\ULon}

\def\hc{c.c.}

\numberwithin{equation}{section}

\allowdisplaybreaks

\allowbreak

%\begin{document}

%%%%%%%%%%%%%%%

\begin{titlepage}
	\thispagestyle{empty}
	\begin{flushright}

%		\hfill{DFPD-2018/TH/xx}
	\end{flushright}
\vspace{35pt}

	\begin{center}
	    { \Large{ 
	  On the F-term problem and quintessence supersymmetry breaking 
	     }}

		\vspace{50pt}

		{\large Fotis~Farakos}

		\vspace{25pt}

		{
			 {\it Dipartimento di Fisica e Astronomia ``Galileo Galilei''\\
			Universit\`a di Padova, Via Marzolo 8, 35131 Padova, Italy }

		\vspace{15pt}

			{\it   INFN, Sezione di Padova \\
		Via Marzolo 8, 35131 Padova, Italy}
            
 		}

		\vspace{40pt}

		{ABSTRACT}
	\end{center}

	\vspace{10pt}

Inspired by the stringy quintessence F-term problem we highlight a generic contribution to the effective moduli masses that arises due to kinetic mixings between the moduli and the quintessence sector. We then proceed to discuss few supergravity toy models that accommodate such effect, and point out possible shortcomings. Interestingly, in the standard 2-derivative supergravity action there is no term to mediate the supersymmetry breaking from the kinetic quintessence sector to the gaugini and generate Majorana masses. Therefore we also propose a 2-derivative supersymmetric invariant that plays exactly this role.

\bigskip

\end{titlepage}

%%%%%%%%%%%%%%%

\vskip 0.5cm

\vspace{0.5cm}

\def\thefootnote{\arabic{footnote}}
\setcounter{footnote}{0}

\baselineskip 6 mm

%%%%%%%%%%%%%%%%%%%%

%\tableofcontents

%%%%%%%%%%%%%%%%%%%%%

%\newpage

%%%%%%%%%%%%%%%

%\tableofcontents

\section{Introduction}

The existence of de Sitter vacua has been often challenged in string theory \cite{Danielsson:2018ztv,Obied:2018sgi,Andriot:2018wzk} 
and one needs to contemplate on viable alternatives. 
In particular one scenario that has regained attention is the 
so-called quintessence phase \cite{Wetterich:1987fm,Ratra:1987rm,Caldwell:1997ii}. 
Quintessence is essentially nothing but a low-scale inflationary phase, 
and as such, 
quintessence is still plagued with some of the problems of inflation in string theory.

However, apart from the challenges that one faces when embedding the quintessence sector in string theory 
and achieving moduli stabilization, 
there is an extra issue that arises which relates to the mediation of the supersymmetry breaking 
to the observable sector. 
This issue was highlighted recently in \cite{Hebecker:2019csg}, where it was dubbed ``F-term problem'', 
and can be summarized as follows: 
Even though the net supersymmetry breaking scale can be (at best) of order TeV (in order to control loop corrections), 
the supersymmetry mediation scale is large and gives very light superpartner masses for the standard model 
sector.\footnote{A similar issue (but not really of the same nature as the F-term problem) 
also arises in simple supergravity quintessence models (see e.g. \cite{Copeland:2000vh}).} 
In other words, 
even though the net supersymmetry breaking can be TeV, 
one finds the gaugini masses to be of order $10^{-15}$ TeV. 
One could introduce an extra hidden sector (say $X$) that also breaks supersymmetry, 
but this will not solve the problem. 
Indeed, 
let us assume the quintessence dynamics requires a scalar potential $V_{quint}$ and that 
the breaking from $X$ is mediated as usual by a new term in the K\"ahler potential of the form 
\be
\delta K \sim \frac{\alpha_i}{\Lambda^2}  |X|^2 |\Phi^i|^2 \ \to \ \frac{\alpha_i |F^X|^2}{\Lambda^2}  |A^i|^2  \,, 
\ee
where $\alpha_i$ are some parameters that are expected to be of ${\cal O}(1)$ 
and $\Lambda$ is some cut-off (or equivalently a mediation scale). 
As a result, 
to have a considerable contribution to the non-supersymmetric masses 
one will need a large value for $\langle F^X \rangle \sim($TeV$)^2$. 
The latter will then feed into the supergravity scalar potential 
giving an extra contribution proportional to 
\be
\label{F-prob}
\delta V_{SUGRA} \sim \text{SUSY breaking}  \sim |F^X|^2 \gg V_{quint} \,, 
\ee
which in turn leads to an unrealistic late-time cosmological scenario, 
or at least spoils the original quintessence phase that was controlled by $V_{quint}$. 
One could cancel this new contribution to the scalar potential 
by finding another new contribution to the gravitino mass 
(recall that $V_{SUGRA} = -3 m_{3/2}^2$ + \dots), 
but such resolution posses a challenge as discussed in \cite{Hebecker:2019csg} 
because no new contributions to the gravitino mass seem possible from string theory. 
Or if they exist, they are going to be sub-dominant.

It is important to appreciate that the only consistent way to lower the vacuum energy in N=1 supergravity is 
either by lowering the net supersymmetry breaking or by increasing the gravitino mass. 
Other effects will either fall under these two categories or will include ghosts. 
This is what makes the F-term problem a very serious issue for quintessence models. 
Note also that there is no straightforward way to escape from the F-term problem even if one uses 
the huge freedom granted by non-linearly realized supersymmetry \cite{Rocek:1978nb,Casalbuoni:1988xh}. 
Indeed, the supersymmetry breaking sector, 
even if it is a pure goldstino, 
is typically tied to a non-vanishing contribution to the vacuum energy of the form \eqref{F-prob}, 
which according to \cite{Hebecker:2019csg} has to be extremely small. 
Then, technically, 
it is possible to eliminate the superpartners with various $X$-related nilpotency conditions \cite{DallAgata:2016syy}. 
However, because the supersymmetry breaking scale from $F^X$ is too low, 
the goldstino self-interactions remain non-unitary 
even at rather very low energies (energies comparable to $\sqrt{F^X}$) \cite{Casalbuoni:1988sx}. 
This means such a theory will have a very low cut-off (below the TeV scale). 
Alternatively, 
one could consider that it is the net supersymmetry breaking (which is of order TeV) that enters the nilpotency conditions and not only $\sqrt{F^X}$. 
It is not clear how realistic it is to obtain such a scenario from string theory.

In this work we want to take a closer look into the quintessence phase, 
and its supergravity embedding, 
and see if there is some additional contribution to the scalar masses. 
We first recall that the mass splitting between the component fields of a supermultiplet 
is typically controlled by the overall supersymmetry breaking, 
therefore one should consider the contributions from all relevant sectors. 
Because during quintessence 
we are on a background where there exists a scalar with non-vanishing kinetic energy, 
say $\dot \phi \ne 0$, 
then we will have an extra contribution to the supesymmetry breaking. 
The easiest way to convince oneself that this is the case is by noticing 
that 4D N=1 supergravity is equipped with the appropriate goldstino-gravitino mixing term \cite{Wess:1992cp,FVP} 
\be
\partial_m \phi \overline \chi^\phi \overline \sigma^m \sigma^n \overline \psi_m 
\ \to \ \ 
\dot \phi  \overline \chi^\phi \overline \sigma^m \sigma^0 \overline \psi_m \,. 
\ee
Such coupling with the gravitino is the signature of a supersymmetry breaking sector, 
and means that $\dot \phi$ is generically bound to contribute to mass-splittings within a supermultiplet 
or give mass to scalar moduli. 
This also means that the goldstino is not only provided by the multiplet with the non-vanishing 
auxiliary field VEV (e.g. $F^X$) but also gets a contribution from the fermion superpartner $\chi^\phi$ of the quintessence scalar. 
Schematically we conclude that \cite{Giudice:1999am} 
\be
\label{Total-SUSY}
\text{SUSY breaking} \sim \text{``}|F^X|^2\text{''} + \text{``} \dot \phi^2\text{''} \,. 
\ee
Even though we have established that $\dot \phi$ contributes to the supersymmetry breaking, 
we are again faced with a new type of F-term problem because if $\dot \phi$ is too large, 
then it will endanger the quintessence phase simply by breaking the slow-roll conditions. 
This can be also understood by inserting \eqref{Total-SUSY} into \eqref{F-prob}. 
In fact during a slow-roll phase 
\be
\dot \phi^2 \sim \epsilon H^2\, , 
\ee
and therefore the true challenge in these models is to see how realistic a strong mediation term will be. 
In other words, 
we want to see if we can have parametrically large masses (compared to $\epsilon H^2$), 
that are generated now by $\dot \phi \ne 0$ instead of $F^X \ne 0$. 
As we will see the applicability of such mechanism in quintessence is always model dependent, 
and it may happen that in models derived from string theory such effect cannot truly help ameliorate the F-term problem.

The rest of the article is organized as follows. 
In the next section we work with a non-supersymmetric gravitational theory and in the third section we turn to supergravity. 
In both cases we show that the contribution to the mass of the scalars is model dependent, 
and we also discuss few examples where it works and examples where it fails. 
We also see why such mechanism (which is intrinsic in supergravity embeddings) 
may instead pose a catastrophic threat to otherwise healthy quintessence models, depending on the kinetic mixings. 
This effect is already studied within the context of inflationary cosmology, because it can ruin inflation, 
and therefore is dubbed ``geometrical destabilization'' \cite{Renaux-Petel:2015mga,Cicoli:2018ccr,Grocholski:2019mot,Cicoli:2019ulk}. 
In the fourth section we turn to the gaugini masses which are of course also influenced by the F-term problem. 
We find that the term that mediates the breaking from the quintessence sector to the gaugini is not 
present in standard supergravity. 
Then we proceed to construct a new term that does exactly that and we study its properties. 
It seems that such term finds its natural place within the new-minimal formulation of supergravity, 
and in models where the quintessence phase is driven by a real linear multiplet instead of a chiral. 
In the fifth section we give a few concluding remarks and an outlook for future work.

\section{Effective masses from kinetic mixings}

Before turning to supergravity it is very instructive to work with a non-supersymmetric model. 
To this end let us consider quintessence driven for simplicity by a 
real scalar $\tau$ slow-rolling down a run-away potential, 
that is we simply have 
\be
\label{Quint-sc}
e^{-1} {\cal L}_{quint} = -\frac12 R 
- \frac12 k(\tau) \partial \tau \partial \tau 
- V \,. 
\ee 
Here for later convenience we have also included a kinetic function $k(\tau)$. 
On an FLRW background we have 
\be
ds^2 = - dt^2 + a(t)^2 d\vec{x}^2 \, , 
\ee
where $a$ is the scale factor and the Hubble scale is $H = \dot a / a$. 
The scalar equation of motion and the Friedmann equation then read 
\be 
k(\tau) \ddot \tau + 3 H k(\tau) \dot \tau + \frac12 k'(\tau) \dot \tau^2 + V'(\tau) = 0 \ , \quad  
\frac12 k(\tau) \dot \tau^2 + V = 3 H^2 \, , 
\ee
where the dot refers to time derivative. 
One then needs to achieve slow-roll, just as in standard inflation, 
which requires 
\be
\epsilon = \frac12 \frac{V'^2}{k V^2} \ll 1 \ , \quad  
\eta = \frac{V''}{k V} - \frac12 \frac{k' V'}{k^2 V^2}  \ll 1 \,. 
\ee
Then once the slow-roll conditions are met we have 
\be
\dot \tau \simeq - \frac{V'}{3 H k} \,.  
\ee

Let us assume there is now another scalar in the theory (say $\rho$) 
with canonical kinetic term but 
with a kinetic coupling to the quintessence sector of the form 
\be
k(\tau) \ \text{in \eqref{Quint-sc}} \ \to \ \ \tilde k(\tau) = k(\tau) - \gamma  \rho^2 \, . 
\ee
Note that the scalar $\rho$ can also enter the scalar 
potential $V$ and could also have a mass term due to other effects. 
However such scalar gets an extra contribution to its effective mass 
due to the kinetic coupling 
\be
+ \frac12 e \gamma \, \partial \tau \partial \tau  \rho^2 \ \to \ \ -  \frac12 e \gamma \, \dot \tau^2  \rho^2  \, . 
\ee 
Therefore this new contribution to the mass has the form 
\be
\delta m^2_\rho = \gamma \dot \tau^2 \simeq \gamma \frac{V'^2}{3 V k^2} \simeq \frac{2 \gamma}{3k} \epsilon V \,. 
\ee
As long as slow-roll holds clearly $\epsilon V$ is a small number, 
however the $2 \gamma / 3k$ term can help give a significant positive contribution to the mass, 
or reduce it depending on the sign of $\gamma$. 
In addition, 
we can assume that during the quintessence phase $\rho |_{quint} = 0 = \dot \rho |_{quint} $ 
such that $\tilde k |_{quint} = k |_{quint}$ 
and so the quintessence phase is left intact. 
Note that quintessence has a crucial difference compared with inflation: 
During quintessence it is sufficient to have 
\be
\epsilon \lesssim 1 \, , 
\ee
and so if we take roughly 
\be
V \sim 10^{-120} \ , \quad 
\frac{k}{\gamma} \sim 10^{-90} \, , 
\ee
we get a mass contribution of order (we restore momentarily $M_P$) 
\be
\delta m_\rho \sim 10^{-15} M_P  \,. 
\ee
At first sight such small values for $k/\gamma$ seem to require significant tuning 
and may seem unrealistic. 
Therefore as we said earlier the challenge is to see if such values could be achieved in a realistic model.

A model independent discussion could only get us this far, 
so let us now work with a simple example. 
We set 
\be
\label{gen-ex}
k(\tau) = \frac{1}{\tau^2} \ , \quad V(\tau) = V_0 \tau^{-1/2} \,, 
\ee
and we remind the reader that we are always working with Planck units, 
i.e. $M_P=1$, 
unless otherwise noted. 
Then we can recast the scalar $\tau$ into a form with canonical kinetic terms, 
which means we set 
\be
\phi = ln \tau \, , 
\ee
which brings the scalar potential for the canonical field $\phi$ to the form 
\be
V = V_0 \, e^{- \phi / 2} \,. 
\ee
This scalar potential is typical for quintessence and one can find that the time dependence is 
\be
\phi \simeq 4 \, ln \left( \frac{\sqrt{V_0}}{8 \sqrt 3} t \right) \ , \quad \dot \phi \simeq \frac{H}{2} \simeq \frac{4}{t} \,.  
\ee
Now we can directly evaluate the contribution to the effective mass 
\be
\label{EX-DM}
\delta m^2_\rho = \gamma \dot \tau^2 \simeq \gamma e^{2 \phi} \dot \phi^2 \simeq \frac{\gamma V_0^4}{(18)^2} H^{-6} \,.  
\ee
Taking into account that $V_0^2 H^{-3}$ can take a variety of values 
we see that the mass contribution due to the kinetic mixing can be arbitrarily large 
and can strongly stabilize the scalar as long as $\gamma$ is positive. 
In addition the quintessence phase is not broken because the net kinetic energy is still sub-dominant, 
i.e. we have 
\be
 \frac12 k(\tau) \dot \tau^2 \simeq \epsilon H^2 \sim  \frac{H^2}{8} \ll 3 H^2 \, , 
\ee
which means our slow-roll approximations are of course valid.

It is interesting to get some feedback from the swampland conjectures to see if such hierarchy 
has at least a remote chance of being generated in string theory. 
The distance conjecture for an inflating theory takes roughly the form \cite{Ooguri:2006in,Scalisi:2018eaz} 
\be
\Delta \phi \lesssim \frac{1}{\lambda} log \, \frac{1}{\Lambda_{UV}} \,, 
\ee
where $\Lambda_{UV}$ is some high energy cut-off scale associated to quantum gravity 
and $\lambda$ is generically assumed to be of order one. 
This restricts the range of $\phi$ but still gives a window 
that allows the $V_0^2 H^{-3}$ fraction in \eqref{EX-DM} to take a variety of different values. 
Let us assume 
\be
\Lambda_{UV} \sim \text{TeV} \sim 10^{-15} \ , \quad  \lambda \sim \frac{1}{10} \, , 
\ee
which give 
\be
\Delta \phi  \lesssim 150 \,. 
\ee
Note here that we took the TeV scale (the scale of the superpartners) as the UV cut-off and we set for $\lambda$ 
to be one order of magnitude smaller than the generic expectation. 
Then we can have 
\be
\label{num-1}
\phi \sim 90 \ , \quad V_0 \sim 10^{-100} \, , 
\ee
which give 
\be
\label{num-2}
H \sim 10^{-60} \ , \quad \delta m_\rho \sim \sqrt{\gamma} \times 10^{-15} \,. 
\ee
We see that if $\gamma$ is of order one then such a scenario is marginally within the limits set by the swampland. 
Clearly, 
by changing the behavior of $\gamma$ we can relax the restrictions from the distance conjecture even more.

In the example we discussed it seems possible that the mass due to kinetic mixing will have a significant positive contribution. 
This may be true for the specific model we studied but it is strongly a model-dependent result. 
Indeed, 
if for example $\gamma$ is not a constant, 
and if it is instead given by $\gamma \sim k(\tau)$ then the contribution to the mass of $\rho$ would be 
proportional to $H^2$ and so it would be insignificant. 
Clearly different choices of $\gamma$ lead to masses of different magnitude. 
Another situation is to have $\gamma<0$ in which case the mass 
from the kinetic mixing would tend to make the scalar a tachyon. 
We also note that this effect is typically ignored during inflation because in the deep inflating regime we have $\epsilon \ll 1$ 
and so it is very hard to generate significant contributions to the masses of the other scalars (due to this mechanism). 
For situations where this effect can endanger inflation due to the existence of light moduli 
see e.g. \cite{Renaux-Petel:2015mga,Cicoli:2018ccr,Grocholski:2019mot,Cicoli:2019ulk}.

\section{Examples in N=1 supergravity}

The bosonic sector of the Lagrangian for chiral scalar superfields coupled to N=1 supergravity has the form 
\be
e^{-1} {\cal L} = -\frac12 R - K_{i \overline j} \partial A^i \partial \overline A^j - V_{SUGRA} \, , 
\ee
where $V_{SUGRA}$ is the standard scalar potential of supergravity, see e.g. \cite{Wess:1992cp,FVP}, 
and $K$ is the K\"ahler potential. 
The $A^i$ are the complex scalars that belong to the chiral superfields $\Phi^i$. 
Models and discussions for quintessence in supergravity and in string theory can be found for example in  \cite{Binetruy:1998rz,Brax:1999gp,Copeland:2000vh,Hellerman:2001yi,Cicoli:2012tz,Akrami:2017cir,Chiang:2018jdg,Emelin:2018igk,Farakos:2019ajx,Ferrara:2019tmu,Farakos:2020jbx,Bento:2020fxj}, 
a general review focused on string theory models can be found in \cite{Cicoli:2018kdo} (and in \cite{Hebecker:2019csg}), 
and some alternative proposals in string theory can be found in \cite{Bento:2020fxj,Olguin-Tejo:2018pfq,DallAgata:2019yrr}. 
Therefore we will not commit ourselves here to finding an appropriate superpotential $P(\Phi^i)$ that gives to $V_{SUGRA}$ 
the required quintessence form. 
Rather, we will investigate what types of K\"ahler potentials can give rise to a significant mass to the scalars due to the kinetic mixing, 
assuming we are in a quintessence phase. 
Since the impact of the kinetic mixing on the mass is highly model dependent, 
we will illustrate the various possibilities with few examples.

\subsection*{K\"ahler moduli inspired example}

Let us first give an example where the effect of the quintessence phase 
has a different behavior than the one we saw in the previous section. 
We focus on only two complex scalars $A$ and $T$, 
the latter being the quintessence field. 
We set 
\be
\label{-2K}
K = - 2 \, ln (T + \overline T - A \overline A / 2) \, , 
\ee
where $T$ can be some K\"ahler modulus and $A$ a modulus related for example to the position of 
a D3-brane (see e.g. \cite{Conlon:2005ki,Hebecker:2019csg}). 
Within a complete string theory setup the $K = -2 \dots$ would rather be $K=-3 \dots$ 
(for example if we considered K\"ahler moduli quintessence \cite{Cicoli:2012tz}), 
but since it does not play a significant role we keep the $-2$ so that we match the quintessence 
dynamics with the previous discussions. 
The kinetic terms of $T$ are controlled by 
\be
\label{KTT-ss1}
K_{T \overline T} 
= \frac{2}{(T + \overline T)^2} \left(1 + \frac{A \overline A}{2 (T + \overline T)} + \dots \right)  \,. 
\ee
To study the dynamics let us split the complex scalar field $T$ as 
\be
T = \tau + i \zeta \,, 
\ee
with the fields during the quintessence phase given by  
\be
\zeta|_{quint} = 0 \ , \quad \tau|_{quint}  = \text{slow-rolling scalar}, 
\ee
and 
\be
A|_{quint} = 0 = \dot A|_{quint} \,. 
\ee
Note that the model we have here gives exactly the same form 
for the kinetic function as we had in the previous section $k(\tau) = (1 / 2) \tau^{-2}$. 
It is then convenient to assume that the dynamics of $\tau$ are similar to the ones we found for the working example 
in the previous section (i.e. a scalar potential similar to \eqref{gen-ex}) such that 
\be
\tau \sim V_0^2 H^{-4} \ , \quad \dot \tau \sim V_0^2 H^{-3} \,. 
\ee
We can now evaluate the contribution to the effective mass of $A$ 
due to the kinetic term of $T$ on the quintessence background. 
From \eqref{KTT-ss1} we have 
\be
{\cal L}_{kin \, T} \sim  \frac{|\dot T|^2 |A|^2}{(T + \overline T)^3}  \sim  \frac{\dot \tau^2}{\tau^3} |A|^2 \sim  V_0^{-2} H^6 |A|^2 \, . 
\ee 
To correctly identify the contribution to the mass of $A$ we have 
to take into account also the normalization of its kinetic terms 
\be
{\cal L}_{kin \, A} \sim - \frac{1}{\tau} |\partial A|^2 \sim   - V_0^{-2} H^{4} |\partial A|^2\,. 
\ee
Therefore after canonical normalization we find a very small negative contribution to the effective mass 
\be
\label{m-can}
\delta m^2_{A} \sim - \frac{\dot \tau^2}{\tau^2} \sim - \epsilon \, H^2 \,. 
\ee
If the scalar $A$ is very light this will lead to a destabilization \cite{Renaux-Petel:2015mga,Cicoli:2018ccr,Grocholski:2019mot,Cicoli:2019ulk}, 
but if there is already another positive supersymmetric mass (or not supersymmetric) then there is no issue. 
Note also that the mass is exactly of the same order as the supersymmetry breaking sourced by the quintessence sector, 
that is $\epsilon H^2$.

A different potential can change the impact of such effect and may lead to a significant contribution to the moduli masses. 
Indeed, 
let us assume that we have instead of the scalar potential in \eqref{gen-ex}, 
a scalar potential of the form 
\be
\hat V(\tau) = V_0 \tau^{1/2} \, , 
\ee
but we keep of course the same K\"ahler potential \eqref{-2K}. 
Here the crucial difference is that $\tau$ goes to smaller and smaller values as the quintessence phase proceeds. 
Then we can again recast the scalar $\tau$ into a form with canonical kinetic terms by setting 
\be
\phi = - ln \tau \,. 
\ee
This brings the scalar potential for the canonical field $\phi$ to the form 
\be
\hat V = V_0 e^{- \phi / 2} \,. 
\ee
Then we have $\tau \sim V_0^{-2} H^4$ and thus 
\be
\label{kin-H-2}
{\cal L}_{kin \, T} \sim \frac{\dot \tau^2}{\tau^3} |A|^2 \sim V_0^{2} H^{-2} |A|^2 \, , 
\ee
which may give a considerable negative contribution to the moduli masses and completely ruin quintessence, 
as happens with the geometrical destabilization \cite{Renaux-Petel:2015mga,Cicoli:2018ccr,Grocholski:2019mot,Cicoli:2019ulk}. 
Note however that once canonical normalization is taken into account we will have a mass of the form \eqref{m-can} 
which seems less dangerous. 
Still, the $H^{-2}$ that comes from the kinetic mixing \eqref{kin-H-2} is a huge negative contribution which has to be matched 
by another contribution to the second derivative of the supergravity scalar potential in order to stabilize the scalars.

Note that a crude analysis of the fibre quintessence scenario \cite{Cicoli:2012tz} shows that if we included the D3-brane moduli 
as in \eqref{-2K} and have $T$ as the quintessence scalar then we would get a mass contribution 
of the form $\delta m^2 \sim - \epsilon H^2$. 
This is of course not an immediate threat as long as there is a sector that 
contributes to the moduli masses at least as $H^2$. 
Such contribution is within reach even if the F-term problem persists.

Until now we have seen that the contribution to the moduli masses due to the quintessence sector is 
of order $\epsilon H^2$, and so rather insignificant. 
Needless to say that one can add modifications to the K\"ahler potential \eqref{-2K} such that 
the contribution from quintessence changes and gives instead a huge impact to the masses. 
For example a succinct 
list of the contributions that can enter \eqref{-2K} can be found in \cite{Hebecker:2019csg}. 
However, 
we checked few simple deformations and found that 
the parameters entering such modifications would require a significant amount of tuning to increase considerably the 
contribution to the effective mass. 
Therefore one can wonder if the required modifications, 
and in particular the amount of tuning required, would really exist in string theory. 
We will come back later to this point and give a working example.

\subsection*{General expectations}

As we have seen the quintessence sector seems to give very small contributions to the effective masses of the moduli. 
In search of possible modifications, 
let us now see what are the general expectations we can have in a supergravity theory. 
Let us first assume we have a K\"ahler potential $K(T+\delta T, 
\overline T + \delta \overline T, 
A + \delta A, 
\overline A + \delta \overline A)$ 
with a form such that it can be expanded as follows 
\be
K = \alpha |\delta T|^2 + \beta |\delta A|^2 - \gamma |\delta A|^2 |\delta T|^2 + \dots
\ee
where the dots are simply higher order terms. 
The coefficients $\alpha$, $\beta$ and $\gamma$ are of course nothing but the derivatives of the K\"ahler potential 
with respect to the chiral superfields, e.g. $\alpha = K_{T \overline T}$, $\beta = K_{A \overline A}$, 
and 
\be
\gamma = - K_{T \overline T A \overline A} \, , 
\ee
and they are  in principle field-dependent. 
Then the consistency of the kinetic terms will require 
\be
\alpha(T, \overline T) > 0 \ , \quad \beta(T, \overline T) - \gamma(T, \overline T) |\delta T|^2 > 0 \, ,  
\ee
and we also choose to have $\gamma (T, \overline T)>0$. 
As before, for quintessence we have 
\be
\delta T|_{quint} = \tau \,, 
\ee
which is a real scalar. 
Now the kinetic mixing will induce a mass term to the scalar $A$ and once we also take into account canonical normalization we have 
\be
\label{gen-mass-A} 
\delta m^2_{A} = \frac{\gamma \, \dot \tau^2}{\beta - \gamma \tau^2} \,. 
\ee
For this effect to dominate the mass of the complex scalar $A$ one would essentially ask that $\gamma$ takes parametrically large values. 
However, 
then, 
the denominator due to the canonical normalization would become negative 
signaling that the $A$ scalar would be a ghost. 
In other words, to guarantee that the denominator is positive one has to ask that $\beta > \gamma \tau^2$ 
which can be safely satisfied only when $\gamma$ is not parametrically large.

One can be tempted to take $\beta \gtrsim \gamma \tau^2$ in \eqref{gen-mass-A} 
such that the denominator remains positive but approaches zero, 
in which case the mass becomes arbitrarily large. 
However such large mass should not be attributed to the quintessence supersymmetry breaking per se, 
rather it is due to the singular kinetic term of $A$. 
Indeed, 
one important point we would like to discuss is related to the fermionic and the scalar moduli kinetic terms. 
In principle due to the kinetic mixings we will have terms of the form 
\be
- K_{i\overline j}(\tau) \, \partial A^i \partial \overline A^j 
- i K_{i\overline j}(\tau) \, \overline \chi^{j} \overline \sigma^m D_m(\omega) \chi^i \, , 
\ee
which after the field redefinitions may in any case alter the masses. 
However such terms do not require a specialized discussion here for the following reasons. 
Firstly, 
such terms lead to a rescaling of the fermions and the scalars of the same supermultiplet 
in exactly the same way. 
So as far as the F-term problem is concerned such terms are not so important 
because they do not generate a mass splitting per se. 
Secondly, 
these terms will influence also the moduli masses 
that are generated from the standard scalar potential. 
Therefore, 
if these terms do have a significant impact on the masses, 
then they will also affect the mass splitting coming from the F-term breaking. 
So, again, it is not an effect that should be attributed to the quintessence supersymmetry breaking per se. 
Finally, 
if we want to address the scalar moduli stabilization, 
then such terms should be in any case taken into account when we evaluate the moduli masses, 
and so our discussion does not have something extra to add to that. 
However, we always do have to check in the end the true effective mass of a scalar, 
as we have been doing.

Our general discussion until now seems to imply that we could never make the kinetic mixing give a huge 
positive contribution to the masses in a supergravity setup, and therefore much less in string theory. 
However, 
there is a small detail in the K\"ahler potential that changes completely the behavior of such term. 
In particular we simply have to ask that 
\be
\label{gAA}
K_{A \overline A} = \beta( A , \overline A)  
+ \frac{1}{2} (T-\overline T)^2 \, \gamma (T, \overline T, A , \overline A) \,. 
\ee
Then the canonical mass of the scalar $A$ during a quintessence phase 
(that is  Re$\,\delta T = \tau$ and Im$\,\delta T = 0$),  
will receive instead a contribution of the form 
\be
\delta m^2_{A} = \frac{\gamma(\tau) \, \dot \tau^2}{\beta(\tau)} \,. 
\ee
As a result we see that by making $\gamma$ arbitrarily large, 
which we are now at least technically allowed to do, 
we can generate a parametrically large effective mass for the scalar $A$.

The form of the K\"ahler metric \eqref{gAA} implies that in that sector the real part of $T$ 
has at least some sort of shift symmetry. 
In fact the requirements for shift symmetry may become even stronger when one tries to build realistic 
scalar potentials, i.e. once we introduce a superpotential and check slow-roll, 
stability, etc. 
Note also that the form of \eqref{gAA} guarantees that the quintessence scalar will interact with $A$ 
only when the derivatives are acting on it. 
These properties of the quintessence scalar already remind the properties of the axions. 
In fact it is exactly the derivative couplings of the quintessence sector 
with the other fields that are needed for this mechanism to work. 
We leave a detailed study of the impact of this mechanism on realistic stringy quintessence models 
that are based on axions for a future work. 
Instead now we turn to a working example following the strategy we just outlined.

\subsection*{An ad hoc working example}

Let us now support our previous discussion with a specific example, 
which is inevitably at this point rather ad hoc. 
We set 
\be
K = - 2 \, ln (T + \overline T) + A \overline A 
+ \frac{1}{M^2} (T - \overline T)^2 A \overline A \, .  
\ee 
Then we have the kinetic terms 
\be
e^{-1} {\cal L}_{kin} = 
- \frac{2}{( T + \overline T )^2} \partial T \partial \overline T 
+ \frac{2}{M^2} \partial T \partial \overline T |A|^2 
 - \partial A \partial \overline A 
- \frac{(T - \overline T)^2}{M^2} \partial A \partial \overline A + \dots \,, 
\ee
where the dots stand for terms that are not relevant to us now. 
As before on an FLRW background we have 
\be
+ \frac{2}{M^2} \partial T \partial \overline T |A|^2 
\ \to \ \ 
-2 \frac{|\dot T|^2}{M^2}  |A|^2  \,, 
\ee
which gives rise to a contribution to the effective mass 
\be
\delta m^2_A = 2 \frac{|\dot T|^2}{M^2} \,. 
\ee 
The model we have here gives for the kinetic function the form $k(\tau) = (1 / 2) \tau^{-2}$ and so 
if we worked with a scalar potential similar to \eqref{gen-ex} we would get 
\be
\label{A-mass-big}
\delta m^2_A \sim  \frac{\dot \tau^2}{M^2} \sim \frac{1}{M^2} V_0^4 H^{-6} \,. 
\ee
We see that in this case we do get a significant contribution to the effective mass of the scalar $A$, 
as we anticipated. 
We stress that the mass \eqref{A-mass-big} is truly a product of the supersymmetry breaking 
that is sourced from the kinetic energy of the quintessence scalar  $T$. 
One can explicitly check this on the full supergravity action 
by verifying that the fermion superpartners of $A$ do not get a similar mass term as \eqref{A-mass-big}. 
Note that the kinetic terms of $A$ get an extra contribution due to the term 
$(T - \overline T)^2 \partial A \partial \overline A \sim \zeta^2 \partial A \partial \overline A$.  
However, because in our example here $\zeta=0$ during the quintessence phase, such term vanishes.

\section{Gaugino masses?}

Let us now turn to the gaugino F-term problem. 
Until now we have discussed the impact of the F-term problem on the masses of scalar moduli 
assuming that a similar discussion (at least qualitatively) will hold for sfermions 
(scalars that are superpartners of the standard model fermions). 
However, 
one can wonder what happens to the gaugini, which are also influenced by the F-term problem.  
Indeed, one needs a term of the form 
\be 
\label{X-W2}
\frac{1}{\Lambda} \int d^2 \theta X W^2 \, , 
\ee 
for some cut-off $\Lambda$ (or equivalently, a mediation scale), 
in order to generate standard Majorana masses of the form 
\be 
\frac{F^X}{\Lambda} \lambda^2 \,. 
\ee
As we already discussed, 
in the stringy models either $\Lambda$ is too large or $F^X$ is forced to be small 
and as a result $F^X / \Lambda$ remains very low \cite{Hebecker:2019csg}. 
Moreover, 
we have seen that the quintessence supersymmetry breaking 
may instead give a significant contribution to the scalar masses and we would like to see if it is possible to have a similar effect for the gaugini. 
However the standard 2-derivative supergravity \cite{Wess:1992cp,FVP} is not equipped with a term 
that gives Majorana masses due to the quintessence phase. 
In this section we address exactly this issue. 
First we see that in rigid supersymmetry one can construct 2-derivative terms 
that give rise to gaugini Majorana masses. 
However, these terms cannot be embedded directly in the old-minimal formulation. 
Therefore we turn to an effective field theory approach for their supergravity embedding, 
but we still ask that they give the 2-derivative terms in the rigid limit. 
In the new-minimal formulation instead such terms do exist and describe 2-derivative interactions.

Note that the supergravity theory does instead contain terms that generate fermionic bilinears of the form 
$e^{-1}{\cal L}_{SUGRA} \sim \frac{1}{M_P^2} \, i (K_T  - K_{\overline T} ) \, \dot \tau \, \overline \lambda \overline \sigma^0 \lambda$. 
However such terms do not have the standard Majorana form as they mix the chiralities, 
they decouple in the rigid limit, 
and also equivalent terms further contribute both to the masses of the fermions that belong to the chiral multiplets but also to the gravitino. 
Therefore we avoid invoking such terms, 
and we also notice that they do vanish in our working examples. 
We believe that they deserve a careful study that we leave for future work. 
Here instead we will construct standard Majorana masses.

\subsection*{Tensor multiplets and rigid supersymmetry}

Let us start by introducing a supersymmetric interaction that contains up to two derivatives, 
up to Gaussian auxiliary fields, 
and gives rise to a gaugino mass on a time dependent background. 
This term is exactly the missing piece that helps complete the mediation of the quintessence 
supersymmetry breaking to matter. 
It is more convenient to describe such superspace coupling in terms of a real linear multiplet, 
and then dualize it to derive its form in terms of chiral superfields.

Let us present the superfields we need.\footnote{We use conventions from the book of Wess and Bagger \cite{Wess:1992cp}.} 
A real linear multiplet is defined in superspace as 
\be
L = L^* \ , \quad \overline D^2 L = 0 = D^2 L \,. 
\ee
Its component fields are 
\begin{eqnarray}
L | = a \ , \quad 
D_\alpha L | = \sqrt 2 \chi^L_\alpha  \ , \quad 
- \frac{1}{2} [D_\alpha , \overline D_{\dot \alpha}  ] L |  =  \sigma^m_{\alpha \dot \alpha} H_{m}  \,  , 
\end{eqnarray}
where $H_m$ satisfies the constraint $\partial^m H_m = 0$, 
which means it is the field strength of a real two-form $H_m =  \epsilon_{mnkl} \partial^n B^{kl}$. 
Since we want to couple to the gaugini, 
let us also introduce the standard N=1 gauge multiplet. 
Its component fields reside in 
the gauge invariant chiral superfield 
\be
W_\alpha = - \frac14 \overline D^2 D_\alpha V \,, 
\ee
and have the form 
\begin{equation}
W_\alpha | =  - i \lambda_\alpha \ , \quad  
(D_\alpha W_\beta + D_\beta W_\alpha) | =  2i (\sigma^{nm} \epsilon)_{\alpha \beta} F_{mn} \ , \quad  
D^\alpha W_\alpha | =  - 2 \text{D} \, , 
\end{equation}
where $F_{mn} = \partial_m v_n - \partial_n v_m$. 
Here we will work with an abelian gauge vector but our coupling can be used also for non-abelian. 
Finally the chiral superfield ($\overline D_{\dot \alpha} \Phi = 0$) will be useful later and has the expansion 
\be
\Phi = A + \sqrt 2 \theta \chi^\Phi + \theta^2 F^\Phi \,. 
\ee

The term that we study here and that mediates the supersymmetry breaking is 
\be
\label{med-1}
{\cal L}_{med} = \frac{1}{M^3} \int d^4 \theta L^2 (W^2 + \overline W^2) 
 = - \frac{1}{2M^3} \int d^2 \theta \, W^2 (\overline D L)^2  + c.c. \,, 
\ee
where in the second equality the $\overline D_{\dot \alpha} L$ is allowed because it is a chiral superfield in global supersymmetry. 
Our first job is to verify that this is a two-derivative term and that there are no higher order auxiliary field terms. 
To this end we perform the superspace integral and act with the superspace derivatives to 
bring the mediation term to the form 
\be
\label{med-2}
{\cal L}_{med}  =  \frac{1}{8M^3} D^2W^2 | (\overline D L)^2 | 
+ \frac{1}{4M^3} D^\alpha W^2 | D_\alpha (\overline D L)^2 | 
+ \frac{1}{8M^3} W^2 | D^2 (\overline D L)^2 | + c.c. 
\ee
From this expansion and from the component field definitions 
we see that there is not a single term that will give rise to more than two derivatives 
or to more than two auxiliary fields. 
Moreover, we also see that there are always at least two fermions with no derivatives acting on them, 
and there are never more than one derivatives acting on any field. 
This is crucial because it tells us that once we add this term to 4D N=1 gauged chiral models 
the standard 2-derivative kinetic terms and the scalar potential are left intact. 
We leave the full component field expansion, 
the investigation of the properties of all the terms in \eqref{med-2}, 
and possible generalizations for a future work. 
Here instead we focus on the last term in \eqref{med-2} which gives in components 
\be
\label{med-3}
{\cal L}_{med}  = - \frac{1}{4M^3} \lambda^2 \left( \partial_a a + i H_a \right)^2 + \dots  
\ee
where the dots contain the complex conjugate and other fermionic terms. 
From \eqref{med-3} we see that once the real scalar $a$ or the gauge two-form 
have a time-dependent profile the gaugini will get an effective Majorana mass term.

One can wonder how arbitrary is the coupling we have introduced for the mediation term of the gaugini. 
As we will show now it is rather unique. 
As we have discussed until now we are considering backgrounds where either the real scalar $a$ or the 
gauge two-form get a time dependent profile. 
As a result 
\be
\langle \left( \partial_a a + i H_a \right)^2 \rangle \ne 0 \, , 
\ee
where with the VEV symbol we refer to the dynamical background. 
We then recall that 
\be
\langle D^2 \left(  \overline D L \overline D L \right) \rangle = 2 \langle \left( \partial_a a + i H_a \right)^2 \rangle \ne 0 \,. 
\ee
As a result $\overline D L \overline D L$ is a chiral superfield with a non-vanishing VEV for its $\theta^2$ component. 
These properties suggest that we make the identification 
\be
\label{XDLDL}
X = \overline D_{\dot \alpha} L \, \overline D^{\dot \alpha} L \, . 
\ee
This composite superfield satisfies 
\be
\label{NLX}
\overline D_{\dot \alpha} X = 0 \ , \quad X^2 =0 \ , \quad \langle F^X \rangle \ne 0 \,.
\ee
Therefore we have successfully identified a nilpotent goldstino chiral superfield. 
With this identification the term \eqref{med-1} takes the standard form \eqref{X-W2} that corresponds to the 
mediation of the supersymmetry breaking to the gaugini. 
There is of course further arbitrariness in the exact definition of $X$ in terms of $\overline D L$, 
but now it is clear that it will be related to the arbitrariness in describing the goldstino itself in superspace. 
In the end, it is known that all the formulations are related to each other \cite{Cribiori:2016hdz}. 
Note that the identification \eqref{XDLDL} was also used in \cite{Farakos:2013zsa} for a slightly different setup, 
whereas in \cite{Bagger:1997pi} a similar identification is related to the partial breaking of supersymmetry. 
In \cite{Kuzenko:2017oni} a similar identification is also 
possible as the supersymmetry breaking is sourced by a modified real linear multiplet.

\subsection*{Dualizing to chirals}

As we have seen the new coupling has a very simple form in terms of tensor multiplets, 
but to utilize it in a 4D N=1 supergravity we prefer to recast it in terms of a chiral superfield. 
We follow this path for two reasons: 
First, 
if we couple the term \eqref{med-1} to old-minimal N=1 supergravity a first inspection shows that we are bound to 
find non-Gaussian auxiliary fields and higher derivatives (these two typically go together). 
This problem persists also when we turn to the chiral multiplet description of course, 
as we will see later, so we will in any case treat the new term only as a perturbation in supergravity, 
i.e. we will evaluate it on the existing background. 
Second, since we have been working until now with chiral multiplets it is more convenient to continue in the same framework. 
However, 
we do present the embedding of the term \eqref{med-1} in the new-minimal formulation in a later subsection where we argue that it 
does not contain higher derivatives and the auxiliary fields remain at most Gaussian.

 Now we recast the mediation superspace coupling in terms of chiral superfields. 
To do this we perform the standard chiral-linear duality. 
Notably, 
the fact that we can perform this duality is another indication that there is nothing peculiar with the term \eqref{med-1}. 
We consider the Lagrangian 
\be
\label{d-1}
{\cal L}_D =   - \frac12 \int d^4 \theta L^2 
+ \int d^4 \theta L (\Phi + \overline \Phi) 
 + \frac{1}{M^3} \int d^4 \theta L^2 (W^2 + \overline W^2) \, , 
\ee
where in order to perform the duality we have to pick a kinetic term for $L$. 
Note that due to the term $ \int d^4 \theta L (\Phi + \overline \Phi)$ now $L$ is unconstrained. 
If we integrate out $\Phi$ then $L$ becomes a standard real linear. 
Now we vary the real superfield $L$ instead, 
to get 
\be
L = (\Phi + \overline \Phi)  
+ \frac{2}{M^3} L (W^2 + \overline W^2) \, , 
\ee
which we solve iteratively as 
\be
L = (\Phi + \overline \Phi) \left[ 1 \!+\! \frac{2}{M^3} (W^2 \!+\! \overline W^2) 
\!+\!   \frac{8}{M^3} W^2 \overline W^2 \right] \!. 
\ee
Then we re-insert the solution for $L$ into the dualizer Lagrangian 
and we find the equivalent Lagrangian for $\Phi$. 
The result is 
\be
\label{d-2}
{\cal L}_D =  \frac12 \int d^4 \theta 
(\Phi + \overline \Phi)^2 \left[ 1 +  \frac{2}{M^3} (W^2  +  \overline W^2)  
+ \frac{8}{M^6} W^2 \overline W^2 \right] \,. 
\ee
We see that the dual term has a form that looks like a higher derivative term, 
however, 
in a hidden way it is not. 
This is guaranteed because we derived it from a 2-derivative theory. 
Clearly the two-derivative structure can only be seen after a series of field redefinitions. 
The term that mediates the quintessence supersymmetry breaking to the gaugini is manifest and it has the form 
\be
\frac{1}{M^3} \int d^4 \theta 
(\Phi + \overline \Phi)^2 W^2  +  c.c. \, , 
\ee
which is the form that we will use in the supergravity theory. 
Interestingly we see that there is again a hidden shift symmetry in this coupling, 
reminding the mediation of the breaking to the scalar moduli. 
Notice also that if 
\be
\langle A + \overline A \rangle = 0 \, , 
\ee
then all the higher order terms in \eqref{d-2} will in any case not contribute to the kinetic terms that would arise from the 
superspace derivatives acting on $W^2$ or its complex conjugate. 
Therefore \eqref{d-2} can be also treated as an effective perturbative expansion around the tree-level interactions, 
and this is in fact how we will treat it in supergravity.

\subsection*{Coupling to supergravity}

We can now introduce the supergravity term that can mediate the quintessence supersymmetry breaking to the gaugini. 
To be compatible with our working example in the previous section we will use the chiral superfield $T$ for the quintessence sector, 
and we also preserve the same form for the mild shift symmetry. 
To this end we will work with the term 
\be 
\label{med-1-sg} 
{\cal L}_{med}  = \frac{1}{M^3} \int d^4 \theta E \, 
(T - \overline T)^2 W^2  +  c.c. \, , 
\ee
and assume we couple it to a standard gauged 4D N=1 supergravity model. 
We can conceptually think of having included also the higher order contribution $1/M^6$ of \eqref{d-2} 
so that we get the correct mediation term in the rigid limit, 
but this will not alter our discussion here in supergravity. 
First let us recast \eqref{med-1-sg} in a more familiar form, 
that is 
\be
\label{med-sugra}
\begin{aligned}
{\cal L}_{med}  = & - \frac{1}{4M^3} \left[ \int d^2 \Theta \, 2 {\cal E} \,  W^2  
\overline{\cal D}^2 (T - \overline T)^2 \right] +  c.c. 
\\
& + \frac{2}{M^3} \left[ \int d^2 \Theta \, 2 {\cal E} \, 
 {\cal R} 
(T - \overline T)^2 W^2 \right] +  c.c. 
\end{aligned}
\ee
We will now treat this term in an effective field theory approach and therefore we will 
assume it only influences the theory in a perturbative way. 
To this end we will keep only the leading contributions that survive once we set the fields to their background values. 
In particular, as before, 
we will assume that the complex scalar field $T$ will be given by 
\be
T = \tau + i \zeta \,, 
\ee
with the fields during the quintessence phase to be 
\be
\zeta|_{quint} = 0 \ , \quad \tau|_{quint}  = \text{slow-rolling scalar}. 
\ee
Then on such background there are no bosonic terms arising from \eqref{med-sugra}, 
and the only terms with two fermions are given by 
\be
\label{med-expand}
e^{-1} {\cal L}_{med}  =  \frac{2}{M^3} \partial \tau \partial \tau \lambda^2 
+ \frac{1}{M^3} \left( F_{mn} F^{mn} - 2 \text{D}^2 - \frac{i}{2} F_{mn} F_{kl} \epsilon^{klmn} \right) (\chi^T)^2 + c.c.  \,, 
\ee
where $\chi^T$ are the fermion superpartners of $T$. 
The fact that the terms $(\chi^T)^2$ are multiplied with D$^2$ 
means they get an extra mass contribution if the breaking is sourced also from the gauge multiplet. 
Then we perform the standard Weyl rescaling 
\be
g_{mn} \to e^{K/3} g_{mn} \, , \ \   e \to e^{2K/3} \,e  \, , \ \  \lambda \to e^{-K/4} \lambda \, ,  
\ee
we give to the scalars $\tau$ their time-dependent profile, 
and thus we find the effective masses of the gaugini (assuming a trivial gauge kinetic function) 
\be
\label{mass-g}
\delta m_{gaugini} = - \frac{4}{M^3} e^{-K/6} \, \dot \tau^2 \, . 
\ee
As we have seen in the previous examples such term can give a significant contribution to the effective mass. 
Note of course that we can also include an arbitrary holomorphic function ${\cal M}(T)$ 
in \eqref{med-1-sg} to give different properties to the gaugino mass during the evolution of the quintessence phase, 
and give to the mass \eqref{mass-g} an extra factor ${\cal M}(\tau)$. 
Such modification will become crucial when one focuses on model building, 
but goes beyond the scope of our work here.

\subsection*{Restrictions on the quintessence superpartner}

Let us also take the opportunity here to explore if the superspace couplings we are discussing can have some 
direct signal to the observable sector, 
and if such effects lead to restrictions. 
As we will see the mediation term \eqref{med-1-sg} will impose a very strict phenomenological requirement: 
the fermion superpartner of the quintessence sector has to decouple from the late-time cosmology.

Let us assume that we have a quintessence phase that, 
as in our working example from the second section, 
leads to 
\be
\label{H^2}
\langle \dot \tau^2 \rangle \sim V_0^4 H^{-6} \sim 10^{-40} M_P^4 \,. 
\ee
Here we used the values from \eqref{num-1} and \eqref{num-2}. 
Then if we want the gaugino mass \eqref{mass-g} to be of order TeV$\sim 10^{-15} M_P$, 
we have to set 
\be
\label{M-ex}
M \sim 10^{-8} M_P \sim 10^{11} \, \text{GeV} \,. 
\ee 
We have assumed $e^{-K/6} \sim 1$ here (and in the rest of this subsection) 
without loss of generality because we could in any case invoke the extra holomorphic function 
${\cal M}(T)$ that can enter \eqref{med-1-sg} to cancel the $e^{-K/6}$ factor in \eqref{mass-g}. 
Now, among the various couplings of \eqref{med-expand}, 
after Weyl rescaling, 
we find the derivative interaction 
\be
\label{FF-ex}
{\cal L}_{med}  =  \frac{1}{M^3} e^{-K/6} \, F^{mn}F_{mn} \, (\overline \chi^T)_{\dot \alpha} (\overline \chi^T)^{\dot \alpha} + \dots 
\ee
between the superpartner of the quintessence scalar and the standard model gauge bosons. 
This interaction however is not describing the canonical fermion superpartner of the quintessence scalar. 
Indeed, 
it is now important to appreciate that for fields that belong to the same multiplet we are bound to have 
\be
e^{-1} {\cal L}_{kin} = - \frac12 k(\tau) (\partial \tau)^2 - \frac12 k(\tau) (\partial \zeta)^2  - i k(\tau) \chi^T \sigma^n D_n \overline \chi^T  \, , 
\ee
which means to find the canonical $\chi^T$ interactions we have to rescale \eqref{FF-ex} with $k(\tau)$. 
Then from the consistency of the slow-roll phase one expects to have 
\be
k(\tau) \sim 10^{-80} \, , 
\ee
which means we have to redefine the fermion $\chi^T$ as 
\be
\chi^T \sim 10^{40} \, \chi^T|_{norm} \,. 
\ee
Therefore the true suppression of the higher dimensional operator  \eqref{FF-ex}  in terms of canonically normalized fields 
is very low, 
and instead of $M$, 
it is 
\be
\tilde M \sim 10^{-15} \text{GeV} \,. 
\ee
Such strong interactions of $\chi^T$ with the standard model gauge bosons are in sharp contradiction with the standard model phenomenology. 
However, 
the coupling \eqref{med-expand} itself gives us the answer to this apparent phenomenological shortcoming, 
because it can give to $\chi^T$ a large mass. 
From \eqref{med-expand}, 
once we integrate out the auxiliary field D, 
we also find a term of the form  
\be
\label{DD-ex}
{\cal L}_{med}  = - \frac{2}{M^3} e^{-K/6} \, \mathbb{D}^2 \, (\overline \chi^T)_{\dot \alpha} (\overline \chi^T)^{\dot \alpha} + \dots 
\ee
where now $\mathbb{D}$ is the Killing potential related to the gauging, 
and it is a moduli-dependent function. 
Now, 
it is realistic to assume that due to some shift in the VEVs of the scalar moduli the function $\mathbb{D}$ also gets a VEV, 
that should be of course very small such that the quintessence phase is not threatened. 
For example we could have 
\be
\langle \mathbb{D} \rangle \sim 10^{-62} M_P^2 \,. 
\ee
Then the contribution to the effective mass of the canonically normalized $\chi^T$ will be 
\be
\delta m_{\chi^T} \sim 10^{-10} M_P \sim 10^{9} \, \text{GeV} \,. 
\ee
Therefore the fermion superpartner of the quintessence scalar will be very heavy and essentially 
decouple from the late-time cosmological phase. 
In other words the coupling \eqref{FF-ex} is in fact not part of the low energy effective field theory. 
This also means that the supersymmetry breaking in the low-energy effective field theory 
becomes explicit and not spontaneous.

We can also check what type of interactions are generated once the heavy fermion is integrated out. 
For example, 
if we do not go to the unitary gauge, we have a mixing with the gravitino of the form 
$k(\tau) \dot \tau \, \overline \chi^T \overline \sigma^m \sigma^0 \overline \psi_m$. 
Then once we integrate out $\chi^T$ we find 
$\overline \chi^T \sim k(\tau) \dot \tau \, \overline \sigma^m \sigma^0 \overline \psi_m M^3 / \mathbb{D}^2$. 
Once we insert this back into \eqref{FF-ex} we find an effective interaction of the form 
$F_{mn}^2 \psi_l^2 /M^3$ 
which does not pose a threat to the low energy theory as it is suppressed by the high energy cut-off $M$. 
A complete study of the low energy effective theory after integrating out $\chi^T$ is left for future work.

\subsection*{New-minimal supergravity?}

Until now we have focused on the old-minimal formulation of supergravity. 
We would like now to contemplate on what could be different if one uses the new-minimal formulation 
instead (for new-minimal supergravity see e.g. \cite{Ferrara:1988qxa,Ferrara:2018dyt}). 
The duality between the two formulations has been established in \cite{Ferrara:1983dh} 
but we will see here that some interesting simplifications take place if we discuss the gaugino 
mediation term directly in the new-minimal setup. 
Notably, the duality between new-minimal and old-minimal has been proven in \cite{Ferrara:1983dh} 
only in the presence of chiral and vector multiplets and without higher derivative terms. 
Instead from \eqref{d-2} we see that the chiral superfield version of the gaugino mediation term resembles a higher derivative term, 
which may be the obstruction to performing the full duality. 
Such obstruction would mean that the new-minimal supergravity version of the gaugino mediation term 
could not be described within old-minimal supergravity.

In the new-minimal formulation of supergravity there is a gauged R-symmetry and so one requires 
the chiral superfields that enter the chiral integral $\int d^2 \theta \, {\cal E}$ to have overall chiral weight equal to n=1. 
The chiral weights of the ingredients we will need for our discussion are 
\be
n(L) = 0 \ , \quad  n(W^2) = 1 \ , \quad  n({\cal M}(\Phi^i)) = -1 \, , 
\ee
where $L$ is the real linear superfield, $W_\alpha$ is the standard chiral superfield associated to the gauge theory, 
and ${\cal M}(\Phi^i)$ is a holomorphic function of the chiral superfields $\Phi^i$. 
The latter are allowed to have arbitrary chiral weight. 
Now, 
once we act with a new-minimal superspace derivative on the real linear superfield we get a weight 
\be
n( \overline{\cal D}_{\dot \alpha} L) = \frac12 \,.  
\ee
The advantage of new-minimal supergravity is that $\overline{\cal D}_{\dot \alpha} L$ is also a bona fide chiral multiplet, 
just as in rigid supersymmetry, 
therefore we have 
\be
\overline{\cal D}^2 L = 0 
\ , \quad 
\overline{\cal D}_{\dot \beta} \left( \overline{\cal D}_{\dot \alpha} L \, \overline{\cal D}^{\dot \alpha} L \right) = 0 \,. 
\ee
As a result we are allowed to introduce a term of the form 
\be
\label{new-min}
{\cal L}_{new} = \int d^2 \theta \, {\cal E} \, W^2 \, {\cal M}(\Phi^i) \, \overline{\cal D}_{\dot \alpha} L \overline{\cal D}^{\dot \alpha} L + c.c. 
\ee
The first thing we notice is that, 
in contrast to old-minimal supergravity, 
the term \eqref{new-min} is in fact K\"ahler invariant. 
This happens because in new-minimal supergravity K\"ahler invariance has the same form as in 
the rigid theory, 
that is (for $H$ chiral) we have 
\be
\int d^4 \theta E \tilde K \ \to \ \  \int d^4 \theta E (\tilde K + H + \overline H) =  \int d^4 \theta E \tilde K \, , 
\ee
up to boundary terms of course. 
Here $\tilde K$ is a real function that relates to the K\"ahler potential of standard supergravity, 
that is $K$, 
but is not restricted to have $det \tilde K_{i \overline j}>0$ \cite{Ferrara:1988qxa}. 
As is shown in \cite{Ferrara:1988qxa} the K\"ahler transformation of $K$ does indeed correspond to the transformation 
$\tilde K \to \tilde K + H + \overline H$ of $\tilde K$. 
As a result K\"ahler invariance is not related to super-Weyl transformations here. 
The second observation is that one could identify ${\cal M}(\Phi^i)$ with $P(\Phi^i)^{-1}$, 
where $P(\Phi^i)$ is the superpotential, 
as long as the gravitino mass is non-vanishing on the background. 
Finally we see that on dimensional grounds we need 
\be
[{\cal M}(\Phi^i)] = - 3 \,. 
\ee
Now let us turn to the component field analysis. 
The chiral density in new-minimal supergravity is 
\be
\label{2Ep}
{\cal{E}}=e \left\{ 1+ i\theta \sigma^{a} \overline{\psi}_a 
-\theta \theta\Big{(} \overline{\psi}_a\overline{\sigma}^{ab}\overline{\psi}_b 
\Big{)} \right\} \, . 
\ee
Then one can see that expanding \eqref{new-min} in components gives a result equivalent to \eqref{med-2}, 
and so will not give rise to terms with more than two auxiliary fields or with higher derivatives. 
The properties of the ingredients involved in such expansion can be found for example in \cite{Ferrara:1988qxa,FVP,Ferrara:2018dyt}. 
In particular, 
due to the form of \eqref{2Ep} 
the two-fermi terms have again the form \eqref{med-2} and give rise to a Majorana mass for the gaugino 
on a quintessence background 
\be
\label{new-min-comp}
{\cal L}_{new} = e \frac{{\cal M}(A^i)}{2} \lambda^2 \left( \partial_a a + i H_a \right)^2 + \dots  
\ee
Because again of the structure of the new-minimal supergravity we also notice that 
we can do the identification 
\be
\label{sugra-X}
X = \overline{\cal D}_{\dot \alpha} L \, \overline{\cal D}^{\dot \alpha} L \, . 
\ee
This is a nilpotent chiral superfield with the properties 
\be
\label{NLX-nm}
\overline{\cal D}_{\dot \alpha} X = 0 \ , \quad X^2 =0 \ , \quad \langle F^X \rangle \ne 0 \,. 
\ee
As a result, 
the term \eqref{new-min} is in fact the standard term describing the mediation 
of the supersymmetry breaking to the gaugini also in supergravity, 
namely 
\be
\label{X-sugra-med}
\int d^2 \theta \, {\cal E} \,{\cal M}(\Phi^i) \, X \, W^2 + c.c. 
\ee
From here we can get another indication 
why in the new-minimal formulation the 
quintessence-gaugini 
mediation term is not expected to give rise to higher derivatives. 
We can keep \eqref{X-sugra-med} with $X$ un-restricted, 
and impose that $X= \overline{\cal D}_{\dot \alpha} L \overline{\cal D}^{\dot \alpha} L$ via a term of the form 
\be
\label{dual-sugra}
\int d^2 \theta {\cal E} \, Z X + 2 \int d^4 \theta E \, Z L^2 + c.c. \, , 
\ee
where $Z$ is a chiral Lagrange multiplier of vanishing chiral weight. 
From \eqref{dual-sugra} once we integrate out $Z$ we get \eqref{sugra-X}. 
The important observation now is that \eqref{X-sugra-med} and \eqref{dual-sugra} are terms 
that in principle belong to the standard 2-derivative supergravity, 
which means they are not expected to include higher derivatives or higher order auxiliary fields. 
We leave a careful study of the full component expansion of \eqref{new-min} for a future work, 
where one should also couple to a simple new-minimal supergravity model, 
integrate out the auxiliary fields and study the dynamics.

\section{Discussion and outlook}

There has been a renewed interest in the study of quintessence models within string theory and supergravity. 
This interest has been sparked from the difficulty to identify controlled de Sitter vacua in string theory, 
and from the various swampland conjectures that restrict de Sitter directly \cite{Obied:2018sgi,Andriot:2018wzk,Dvali:2018fqu,Denef:2018etk,Garg:2018reu,Andriot:2018ept,
Ooguri:2018wrx,Hebecker:2018vxz,Banlaki:2018ayh,Dasgupta:2018rtp,Andriot:2018mav,Andriot:2019wrs,
Agrawal:2018own,Kehagias:2018uem,Andriot:2020lea,Bernardo:2020lar} 
or indirectly \cite{Lanza:2020qmt,Bedroya:2020rac,Farakos:2020wfc,Cribiori:2020use}.

In this work we investigated the impact of the quintessence phase on the moduli stabilization 
and the induced mass splitting within supermultiplets. 
We established that there is indeed an additional contribution to the net supersymmetry breaking 
that arises due to kinetic mixings 
and that it my help in addressing the F-term problem \cite{Hebecker:2019csg}. 
The effect of such mixing can go either way: 
in some cases it gives a significant positive mass and helps strongly stabilize the moduli, 
in other cases it may be innocuous, 
and in some instances it may lead to huge tachyonic masses for the moduli and spoil 
quintessence all together. 
Therefore one would have to investigate the impact of these terms independently in each string theory quintessence model. 
Here instead we have only illustrated the various possibilities with some simple examples in 4D N=1 supergravity. 
In addition, 
even if the F-term problem is resolved by some other mechanism, 
the couplings that we discussed here may in any case play a role in the superpartner masses.

We have also presented a specific superspace term that can induce Majorana 
gaugini masses on a quintessence background and studied few of its properties. 
We have seen that in the new-minimal formulation such term takes a very simple form 
and we have argued that it does not lead to higher derivatives 
(or higher order auxiliary field equations). 
It would be even more surprising if such a completely new term also exists in old-minimal supergravity 
in a way that does not give rise to higher derivatives. 
Instead, our strategy here was to identify this term in the rigid limit 
and then we used it as a perturbation in old-minimal supergravity. 
However, 
since the new-minimal formulation of supergravity can consistently accommodate such term, 
one important future direction is to either perform the duality from new-minimal to old-minimal, 
or otherwise study quintessence directly within new-minimal supergravity. 
Massive vector multiplets may offer an interesting framework to construct such models 
and mediate the quintessence supersymmetry breaking \cite{VanProeyen:1979ks,Farakos:2013cqa,Ferrara:2013rsa,Aldabergenov:2016dcu}.

Finally, 
we studied the couplings of the quintessence superpartner with matter 
in a setup where the quintessence supersymmetry breaking generates TeV gaugini masses. 
We found that generically it will have significant interactions with the standard model gauge sector, 
but, 
it will also generically receive a large mass that can be pushed up to the cut-off, 
and so it will decouple from the low energy theory. 
A careful analysis of the resulting low energy theory and its phenomenological implications is left for future work.

\section*{Acknowledgments} 
I would like to thank A. Hebecker and A. Kehagias for very helpful correspondence and N. Cribiori for discussion. 
This work is supported by the STARS grant SUGRA-MAX.


\begin{thebibliography}{} %\thinspace




%\cite{Danielsson:2018ztv}
\bibitem{Danielsson:2018ztv}
U.~H.~Danielsson and T.~Van Riet, 
``What if string theory has no de Sitter vacua?,''
Int. J. Mod. Phys. D \textbf{27}, no.12, 1830007 (2018) 
%doi:10.1142/S0218271818300070
[arXiv:1804.01120 [hep-th]]. 




%\cite{Obied:2018sgi}
\bibitem{Obied:2018sgi}
G.~Obied, H.~Ooguri, L.~Spodyneiko and C.~Vafa,
``De Sitter Space and the Swampland,''
[arXiv:1806.08362 [hep-th]].


%\cite{Andriot:2018wzk}
\bibitem{Andriot:2018wzk}
D.~Andriot,
``On the de Sitter swampland criterion,''
Phys. Lett. B \textbf{785}, 570-573 (2018)
%doi:10.1016/j.physletb.2018.09.022
[arXiv:1806.10999 [hep-th]].





%\cite{Wetterich:1987fm}
\bibitem{Wetterich:1987fm} 
  C.~Wetterich,
  ``Cosmology and the Fate of Dilatation Symmetry,''
  Nucl.\ Phys.\ B {\bf 302}, 668 (1988)
%  doi:10.1016/0550-3213(88)90193-9
  [arXiv:1711.03844 [hep-th]].
  %%CITATION = doi:10.1016/0550-3213(88)90193-9;%%

%\cite{Ratra:1987rm}
\bibitem{Ratra:1987rm} 
  B.~Ratra and P.~J.~E.~Peebles,
  ``Cosmological Consequences of a Rolling Homogeneous Scalar Field,''
  Phys.\ Rev.\ D {\bf 37}, 3406 (1988).
%  doi:10.1103/PhysRevD.37.3406
  %%CITATION = doi:10.1103/PhysRevD.37.3406;%%


%\cite{Caldwell:1997ii}
\bibitem{Caldwell:1997ii} 
  R.~R.~Caldwell, R.~Dave and P.~J.~Steinhardt,
  ``Cosmological imprint of an energy component with general equation of state,''
  Phys.\ Rev.\ Lett.\  {\bf 80}, 1582 (1998)
%  doi:10.1103/PhysRevLett.80.1582
  [astro-ph/9708069].
  %%CITATION = doi:10.1103/PhysRevLett.80.1582;%%




%\cite{Hebecker:2019csg}
\bibitem{Hebecker:2019csg}
A.~Hebecker, T.~Skrzypek and M.~Wittner,
``The $F$-term Problem and other Challenges of Stringy Quintessence,''
JHEP \textbf{11}, 134 (2019)
%doi:10.1007/JHEP11(2019)134
[arXiv:1909.08625 [hep-th]].




 
%\cite{Copeland:2000vh}
\bibitem{Copeland:2000vh} 
  E.~J.~Copeland, N.~J.~Nunes and F.~Rosati,
  ``Quintessence models in supergravity,''
  Phys.\ Rev.\ D {\bf 62}, 123503 (2000)
%  doi:10.1103/PhysRevD.62.123503
  [hep-ph/0005222].
  %%CITATION = doi:10.1103/PhysRevD.62.123503;%% 
 




%\cite{Rocek:1978nb}
\bibitem{Rocek:1978nb}
M.~Rocek,
``Linearizing the Volkov-Akulov Model,''
Phys. Rev. Lett. \textbf{41}, 451-453 (1978)
%doi:10.1103/PhysRevLett.41.451


%\cite{Casalbuoni:1988xh}
\bibitem{Casalbuoni:1988xh}
R.~Casalbuoni, S.~De Curtis, D.~Dominici, F.~Feruglio and R.~Gatto,
``Nonlinear Realization of Supersymmetry Algebra From Supersymmetric Constraint,''
Phys. Lett. B \textbf{220}, 569-575 (1989)
%doi:10.1016/0370-2693(89)90788-0


%\cite{DallAgata:2016syy}
\bibitem{DallAgata:2016syy}
G.~Dall'Agata, E.~Dudas and F.~Farakos,
``On the origin of constrained superfields,''
JHEP \textbf{05}, 041 (2016)
%doi:10.1007/JHEP05(2016)041
[arXiv:1603.03416 [hep-th]].




%\cite{Casalbuoni:1988sx}
\bibitem{Casalbuoni:1988sx}
R.~Casalbuoni, S.~De Curtis, D.~Dominici, F.~Feruglio and R.~Gatto,
``WHEN DOES SUPERGRAVITY BECOME STRONG?,''
Phys. Lett. B \textbf{216}, 325 (1989)
[erratum: Phys. Lett. B \textbf{229}, 439 (1989)]
%doi:10.1016/0370-2693(89)91123-4





%\cite{Wess:1992cp}
\bibitem{Wess:1992cp} 
  J.~Wess and J.~Bagger,
  ``Supersymmetry and supergravity,''
  Princeton, USA: Univ. Pr. (1992) 259 p
  %6 citations counted in INSPIRE as of 08 Nov 2013


  
\bibitem{FVP} 
  D.~Z.~Freedman and A.~Van Proeyen,
  ``Supergravity,'' 
  Cambridge University Press (2012). 
  %%CITATION = INSPIRE-1123253;%%
  
  
  
%\cite{Giudice:1999am}
\bibitem{Giudice:1999am}
G.~F.~Giudice, A.~Riotto and I.~Tkachev,
``Thermal and nonthermal production of gravitinos in the early universe,''
JHEP \textbf{11}, 036 (1999)
%doi:10.1088/1126-6708/1999/11/036
[arXiv:hep-ph/9911302 [hep-ph]]. 
  
  
  
  
%\cite{Renaux-Petel:2015mga}
\bibitem{Renaux-Petel:2015mga}
S.~Renaux-Petel and K.~Turzy\'nski,
``Geometrical Destabilization of Inflation,''
Phys. Rev. Lett. \textbf{117}, no.14, 141301 (2016)
%doi:10.1103/PhysRevLett.117.141301
[arXiv:1510.01281 [astro-ph.CO]].  


%\cite{Cicoli:2018ccr}
\bibitem{Cicoli:2018ccr}
M.~Cicoli, V.~Guidetti, F.~G.~Pedro and G.~P.~Vacca,
``A geometrical instability for ultra-light fields during inflation?,''
JCAP \textbf{12}, 037 (2018)
%doi:10.1088/1475-7516/2018/12/037
[arXiv:1807.03818 [hep-th]].


%\cite{Grocholski:2019mot}
\bibitem{Grocholski:2019mot}
O.~Grocholski, M.~Kalinowski, M.~Kolanowski, S.~Renaux-Petel, K.~Turzy\'nski and V.~Vennin,
``On backreaction effects in geometrical destabilisation of inflation,''
JCAP \textbf{05}, 008 (2019)
%doi:10.1088/1475-7516/2019/05/008
[arXiv:1901.10468 [astro-ph.CO]].




%\cite{Cicoli:2019ulk}
\bibitem{Cicoli:2019ulk}
M.~Cicoli, V.~Guidetti and F.~G.~Pedro,
``Geometrical Destabilisation of Ultra-Light Axions in String Inflation,''
JCAP \textbf{05}, 046 (2019)
%doi:10.1088/1475-7516/2019/05/046
[arXiv:1903.01497 [hep-th]].






%%%%%



%\cite{Ooguri:2006in}
\bibitem{Ooguri:2006in}
H.~Ooguri and C.~Vafa,
``On the Geometry of the String Landscape and the Swampland,''
Nucl. Phys. B \textbf{766}, 21-33 (2007)
%doi:10.1016/j.nuclphysb.2006.10.033
[arXiv:hep-th/0605264 [hep-th]].




%\cite{Scalisi:2018eaz}
\bibitem{Scalisi:2018eaz}
M.~Scalisi and I.~Valenzuela,
``Swampland distance conjecture, inflation and $\alpha$-attractors,''
JHEP \textbf{08}, 160 (2019)
%doi:10.1007/JHEP08(2019)160
[arXiv:1812.07558 [hep-th]].




%%%%%%%%




 
 %\cite{Binetruy:1998rz}
\bibitem{Binetruy:1998rz}
P.~Binetruy,
``Models of dynamical supersymmetry breaking and quintessence,''
Phys. Rev. D \textbf{60}, 063502 (1999)
%doi:10.1103/PhysRevD.60.063502
[arXiv:hep-ph/9810553 [hep-ph]].
 
 


 %\cite{Brax:1999gp}
\bibitem{Brax:1999gp} 
  P.~Brax and J.~Martin,
  ``Quintessence and supergravity,''
  Phys.\ Lett.\ B {\bf 468}, 40 (1999)
%  doi:10.1016/S0370-2693(99)01209-5
  [astro-ph/9905040].
  %%CITATION = doi:10.1016/S0370-2693(99)01209-5;%%
 
  
  
% \cite{Hellerman:2001yi}
\bibitem{Hellerman:2001yi} 
  S.~Hellerman, N.~Kaloper and L.~Susskind,
  ``String theory and quintessence,''
  JHEP {\bf 0106}, 003 (2001)
%  doi:10.1088/1126-6708/2001/06/003
  [hep-th/0104180].
  %%CITATION = doi:10.1088/1126-6708/2001/06/003;%%
 


%\cite{Cicoli:2012tz}
\bibitem{Cicoli:2012tz}
M.~Cicoli, F.~G.~Pedro and G.~Tasinato,
``Natural Quintessence in String Theory,''
JCAP \textbf{07}, 044 (2012)
%doi:10.1088/1475-7516/2012/07/044
[arXiv:1203.6655 [hep-th]].




 
 
%\cite{Akrami:2017cir}
\bibitem{Akrami:2017cir} 
  Y.~Akrami, R.~Kallosh, A.~Linde and V.~Vardanyan,
  ``Dark energy, $\alpha$-attractors, and large-scale structure surveys,''
  JCAP {\bf 1806}, no. 06, 041 (2018)
%  doi:10.1088/1475-7516/2018/06/041
  [arXiv:1712.09693 [hep-th]].
  %%CITATION = doi:10.1088/1475-7516/2018/06/041;%% 





%\cite{Chiang:2018jdg}
\bibitem{Chiang:2018jdg}
  C.~I.~Chiang and H.~Murayama,
  ``Building Supergravity Quintessence Model,''
  arXiv:1808.02279 [hep-th].
  %%CITATION = ARXIV:1808.02279;%% 




%\cite{Emelin:2018igk}
\bibitem{Emelin:2018igk}
M.~Emelin and R.~Tatar,
``Axion Hilltops, Kahler Modulus Quintessence and the Swampland Criteria,''
Int. J. Mod. Phys. A \textbf{34}, no.28, 1950164 (2019)
%doi:10.1142/S0217751X19501641
[arXiv:1811.07378 [hep-th]].



%\cite{Farakos:2019ajx}
\bibitem{Farakos:2019ajx}
F.~Farakos,
``Runaway potentials and a massive goldstino,''
Phys. Rev. D \textbf{99}, no.12, 126004 (2019)
%doi:10.1103/PhysRevD.99.126004
[arXiv:1903.07560 [hep-th]].




%\cite{Ferrara:2019tmu}
\bibitem{Ferrara:2019tmu}
S.~Ferrara, M.~Tournoy and A.~Van Proeyen,
``de Sitter Conjectures in $N$=1 Supergravity,''
Fortsch. Phys. \textbf{68}, no.2, 1900107 (2020)
%doi:10.1002/prop.201900107
[arXiv:1912.06626 [hep-th]].


%\cite{Farakos:2020jbx}
\bibitem{Farakos:2020jbx}
F.~Farakos,
``Quintessence from higher curvature supergravity,''
PoS \textbf{CORFU2019}, 135 (2020)
%doi:10.22323/1.376.0135
[arXiv:2003.09366 [hep-th]].



%\cite{Bento:2020fxj}
\bibitem{Bento:2020fxj}
B.~Valeixo Bento, D.~Chakraborty, S.~L.~Parameswaran and I.~Zavala,
``Dark Energy in String Theory,''
PoS \textbf{CORFU2019}, 123 (2020)
%doi:10.22323/1.376.0123
[arXiv:2005.10168 [hep-th]].



%\cite{Cicoli:2018kdo}
\bibitem{Cicoli:2018kdo}
M.~Cicoli, S.~De Alwis, A.~Maharana, F.~Muia and F.~Quevedo,
``De Sitter vs Quintessence in String Theory,''
Fortsch. Phys. \textbf{67}, no.1-2, 1800079 (2019)
%doi:10.1002/prop.201800079
[arXiv:1808.08967 [hep-th]].




%\cite{Olguin-Tejo:2018pfq}
\bibitem{Olguin-Tejo:2018pfq} 
  Y.~Olguin-Trejo, S.~L.~Parameswaran, G.~Tasinato and I.~Zavala,
  ``Runaway Quintessence, Out of the Swampland,''
  JCAP {\bf 1901}, no. 01, 031 (2019)
%  doi:10.1088/1475-7516/2019/01/031
  [arXiv:1810.08634 [hep-th]].
  %%CITATION = doi:10.1088/1475-7516/2019/01/031;%%



%\cite{DallAgata:2019yrr}
\bibitem{DallAgata:2019yrr}
G.~Dall'Agata, S.~Gonz\'alez-Mart\'\i{}n, A.~Papageorgiou and M.~Peloso,
``Warm dark energy,''
JCAP \textbf{08}, 032 (2020)
%doi:10.1088/1475-7516/2020/08/032
[arXiv:1912.09950 [hep-th]].



%\cite{Conlon:2005ki}
\bibitem{Conlon:2005ki}
J.~P.~Conlon, F.~Quevedo and K.~Suruliz,
``Large-volume flux compactifications: Moduli spectrum and D3/D7 soft supersymmetry breaking,''
JHEP \textbf{08}, 007 (2005)
%doi:10.1088/1126-6708/2005/08/007
[arXiv:hep-th/0505076 [hep-th]].




%\cite{Cribiori:2016hdz}
\bibitem{Cribiori:2016hdz}
N.~Cribiori, G.~Dall'Agata and F.~Farakos,
``Interactions of N Goldstini in Superspace,''
Phys. Rev. D \textbf{94}, no.6, 065019 (2016)
%doi:10.1103/PhysRevD.94.065019
[arXiv:1607.01277 [hep-th]].




%\cite{Farakos:2013zsa}
\bibitem{Farakos:2013zsa}
F.~Farakos, S.~Ferrara, A.~Kehagias and M.~Porrati,
``Supersymmetry Breaking by Higher Dimension Operators,''
Nucl. Phys. B \textbf{879}, 348-369 (2014)
%doi:10.1016/j.nuclphysb.2013.12.016
[arXiv:1309.1476 [hep-th]].



%\cite{Bagger:1997pi}
\bibitem{Bagger:1997pi}
J.~Bagger and A.~Galperin,
``The Tensor Goldstone multiplet for partially broken supersymmetry,''
Phys. Lett. B \textbf{412}, 296-300 (1997)
%doi:10.1016/S0370-2693(97)01030-7
[arXiv:hep-th/9707061 [hep-th]].




%\cite{Kuzenko:2017oni}
\bibitem{Kuzenko:2017oni}
S.~M.~Kuzenko,
``Nilpotent ${\cal N}=1$ tensor multiplet,''
JHEP \textbf{04}, 131 (2018)
%doi:10.1007/JHEP04(2018)131
[arXiv:1712.09258 [hep-th]].




%\cite{Ferrara:1988qxa}
\bibitem{Ferrara:1988qxa}
S.~Ferrara and S.~Sabharwal,
``Structure of New Minimal Supergravity,''
Annals Phys. \textbf{189}, 318-351 (1989)
%doi:10.1016/0003-4916(89)90167-X


%\cite{Ferrara:2018dyt}
\bibitem{Ferrara:2018dyt}
S.~Ferrara, M.~Samsonyan, M.~Tournoy and A.~Van Proeyen,
``Comments on rigid and local supercurrents in ${\cal N}=1$ minimal Supergravity,''
Fortsch. Phys. \textbf{66}, no.8-9, 1800049 (2018)
%doi:10.1002/prop.201800049
[arXiv:1805.09228 [hep-th]].




%\cite{Ferrara:1983dh}
\bibitem{Ferrara:1983dh}
S.~Ferrara, L.~Girardello, T.~Kugo and A.~Van Proeyen,
``Relation Between Different Auxiliary Field Formulations of $N=1$ Supergravity Coupled to Matter,''
Nucl. Phys. B \textbf{223}, 191-217 (1983)
%doi:10.1016/0550-3213(83)90101-3




%%%%%%%




%\cite{Agrawal:2018own}
\bibitem{Agrawal:2018own}
  P.~Agrawal, G.~Obied, P.~J.~Steinhardt and C.~Vafa,
  ``On the Cosmological Implications of the String Swampland,''
  Phys.\ Lett.\ B {\bf 784} (2018) 271
%  doi:10.1016/j.physletb.2018.07.040
  [arXiv:1806.09718 [hep-th]].
  %%CITATION = doi:10.1016/j.physletb.2018.07.040;%%




%\cite{Dvali:2018fqu}
\bibitem{Dvali:2018fqu} 
  G.~Dvali and C.~Gomez,
  ``On Exclusion of Positive Cosmological Constant,''
  Fortsch.\ Phys.\  {\bf 67}, no. 1-2, 1800092 (2019)
%  doi:10.1002/prop.201800092
  [arXiv:1806.10877 [hep-th]].
  %%CITATION = doi:10.1002/prop.201800092;%%





%\cite{Garg:2018reu}
\bibitem{Garg:2018reu} 
  S.~K.~Garg and C.~Krishnan,
  ``Bounds on Slow Roll and the de Sitter Swampland,''
  arXiv:1807.05193 [hep-th].
  %%CITATION = ARXIV:1807.05193;%%
  


%\cite{Kehagias:2018uem}
\bibitem{Kehagias:2018uem} 
  A.~Kehagias and A.~Riotto,
  ``A note on Inflation and the Swampland,''
  Fortsch.\ Phys.\  {\bf 66}, no. 10, 1800052 (2018)
%  doi:10.1002/prop.201800052
  [arXiv:1807.05445 [hep-th]].
  %%CITATION = doi:10.1002/prop.201800052;%%


  
%\cite{Denef:2018etk}
\bibitem{Denef:2018etk} 
  F.~Denef, A.~Hebecker and T.~Wrase,
  ``de Sitter swampland conjecture and the Higgs potential,''
  Phys.\ Rev.\ D {\bf 98}, no. 8, 086004 (2018)
%  doi:10.1103/PhysRevD.98.086004
  [arXiv:1807.06581 [hep-th]].
  %%CITATION = doi:10.1103/PhysRevD.98.086004;%%

  



%\cite{Andriot:2018ept}
\bibitem{Andriot:2018ept} 
  D.~Andriot,
  ``New constraints on classical de Sitter: flirting with the swampland,''
  Fortsch.\ Phys.\  {\bf 67}, no. 1-2, 1800103 (2019)
%  doi:10.1002/prop.201800103
  [arXiv:1807.09698 [hep-th]].
  %%CITATION = doi:10.1002/prop.201800103;%%


%\cite{Dasgupta:2018rtp}
\bibitem{Dasgupta:2018rtp}
K.~Dasgupta, M.~Emelin, E.~McDonough and R.~Tatar,
``Quantum Corrections and the de Sitter Swampland Conjecture,''
JHEP \textbf{01}, 145 (2019)
%doi:10.1007/JHEP01(2019)145
[arXiv:1808.07498 [hep-th]].



%\cite{Ooguri:2018wrx}
\bibitem{Ooguri:2018wrx} 
  H.~Ooguri, E.~Palti, G.~Shiu and C.~Vafa,
  ``Distance and de Sitter Conjectures on the Swampland,''
  Phys.\ Lett.\ B {\bf 788}, 180 (2019)
%  doi:10.1016/j.physletb.2018.11.018
  [arXiv:1810.05506 [hep-th]].
  %%CITATION = doi:10.1016/j.physletb.2018.11.018;%%



%\cite{Hebecker:2018vxz}
\bibitem{Hebecker:2018vxz}
A.~Hebecker and T.~Wrase,
``The Asymptotic dS Swampland Conjecture - a Simplified Derivation and a Potential Loophole,''
Fortsch. Phys. \textbf{67}, no.1-2, 1800097 (2019)
%doi:10.1002/prop.201800097
[arXiv:1810.08182 [hep-th]].


%\cite{Banlaki:2018ayh}
\bibitem{Banlaki:2018ayh} 
  A.~Banlaki, A.~Chowdhury, C.~Roupec and T.~Wrase,
  ``Scaling limits of dS vacua and the swampland,''
  arXiv:1811.07880 [hep-th].
  %%CITATION = ARXIV:1811.07880;%%
 

%\cite{Andriot:2018mav}
\bibitem{Andriot:2018mav} 
  D.~Andriot and C.~Roupec,
  ``Further refining the de Sitter swampland conjecture,''
  Fortsch.\ Phys.\  {\bf 67}, no. 1-2, 1800105 (2019)
%  doi:10.1002/prop.201800105
  [arXiv:1811.08889 [hep-th]].
  %%CITATION = doi:10.1002/prop.201800105;%%



%\cite{Andriot:2019wrs}
\bibitem{Andriot:2019wrs}
D.~Andriot,
``Open problems on classical de Sitter solutions,''
Fortsch. Phys. \textbf{67}, no.7, 1900026 (2019)
%doi:10.1002/prop.201900026
[arXiv:1902.10093 [hep-th]].
  


%\cite{Andriot:2020lea}
\bibitem{Andriot:2020lea}
D.~Andriot, N.~Cribiori and D.~Erkinger,
``The web of swampland conjectures and the TCC bound,''
JHEP \textbf{07}, 162 (2020)
%doi:10.1007/JHEP07(2020)162
[arXiv:2004.00030 [hep-th]].



%\cite{Bernardo:2020lar}
\bibitem{Bernardo:2020lar}
H.~Bernardo, S.~Brahma, K.~Dasgupta and R.~Tatar,
``Crisis on Infinite Earths: Short-lived de Sitter Vacua in the String Theory Landscape,''
[arXiv:2009.04504 [hep-th]].


%\cite{Lanza:2020qmt}
\bibitem{Lanza:2020qmt}
S.~Lanza, F.~Marchesano, L.~Martucci and I.~Valenzuela,
``Swampland Conjectures for Strings and Membranes,''
[arXiv:2006.15154 [hep-th]].


%\cite{Bedroya:2020rac}
\bibitem{Bedroya:2020rac}
A.~Bedroya, M.~Montero, C.~Vafa and I.~Valenzuela,
``de Sitter Bubbles and the Swampland,''
[arXiv:2008.07555 [hep-th]].



%\cite{Farakos:2020wfc}
\bibitem{Farakos:2020wfc}
F.~Farakos, A.~Kehagias and N.~Liatsos,
``de Sitter decay through goldstino evaporation,''
[arXiv:2009.03335 [hep-th]].



%\cite{Cribiori:2020use}
\bibitem{Cribiori:2020use}
N.~Cribiori, G.~Dall'Agata and F.~Farakos,
``Weak gravity versus de Sitter,''
[arXiv:2011.06597 [hep-th]].






%\cite{VanProeyen:1979ks}
\bibitem{VanProeyen:1979ks}
A.~Van Proeyen,
``Massive Vector Multiplets in Supergravity,''
Nucl. Phys. B \textbf{162}, 376 (1980)
%doi:10.1016/0550-3213(80)90345-4


%\cite{Farakos:2013cqa}
\bibitem{Farakos:2013cqa}
F.~Farakos, A.~Kehagias and A.~Riotto,
``On the Starobinsky Model of Inflation from Supergravity,''
Nucl. Phys. B \textbf{876}, 187-200 (2013)
%doi:10.1016/j.nuclphysb.2013.08.005
[arXiv:1307.1137 [hep-th]].

%\cite{Ferrara:2013rsa}
\bibitem{Ferrara:2013rsa}
S.~Ferrara, R.~Kallosh, A.~Linde and M.~Porrati,
``Minimal Supergravity Models of Inflation,''
Phys. Rev. D \textbf{88}, no.8, 085038 (2013)
%doi:10.1103/PhysRevD.88.085038
[arXiv:1307.7696 [hep-th]].



%\cite{Aldabergenov:2016dcu}
\bibitem{Aldabergenov:2016dcu}
Y.~Aldabergenov and S.~V.~Ketov,
``SUSY breaking after inflation in supergravity with inflaton in a massive vector supermultiplet,''
Phys. Lett. B \textbf{761}, 115-118 (2016)
%doi:10.1016/j.physletb.2016.08.016
[arXiv:1607.05366 [hep-th]].


\end{thebibliography}
\end{document}